\documentclass[12pt]{iopart}

\usepackage{graphicx,color}

\usepackage{iopams}  
\usepackage{amsthm}
\usepackage{amssymb}
\usepackage{bm}
\usepackage{bbm}

\begin{document}

\newcommand{\beq}{\begin{equation}}
\newcommand{\eeq}{\end{equation}}
\newcommand{\barr}{\begin{eqnarray}}
\newcommand{\earr}{\end{eqnarray}}
\newcommand{\gv}[1]{\ensuremath{\mbox{\boldmath$ #1 $}}} 

\newcommand{\uv}[1]{\ensuremath{\mathbf{\hat{#1}}}} 
\newcommand{\abs}[1]{\left| #1 \right|} 
\newcommand{\avg}[1]{\left< #1 \right>} 
\let\underdot=\d
\renewcommand{\d}[2]{\frac{d #1}{d #2}}
\newcommand{\dd}[2]{\frac{d^2 #1}{d #2^2}} 
\newcommand{\pd}[2]{\frac{\partial #1}{\partial #2}} 

\newcommand{\pdd}[2]{\frac{\partial^2 #1}{\partial #2^2}} 

\newcommand{\pdc}[3]{\left( \frac{\partial #1}{\partial #2}
 \right)_{#3}} 
\newcommand{\ket}[1]{\left| #1 \right>} 
\newcommand{\bra}[1]{\left< #1 \right|} 
\newcommand{\braket}[2]{\left< #1 \vphantom{#2} |  #2 \vphantom{#1} \right>} 
\newcommand{\matrixel}[3]{\left< #1 \vphantom{#2#3} \right|
 #2 \left| #3 \vphantom{#1#2} \right>} 
\newcommand{\grad}[1]{\gv{\nabla} #1} 
\let\divsymb=\div 
\renewcommand{\div}[1]{\gv{\nabla} \cdot #1} 
\newcommand{\curl}[1]{\gv{\nabla} \times #1} 
\let\baraccent=\= 
\renewcommand{\=}[1]{\stackrel{#1}{=}} 

\newenvironment{sistema}
{\left\lbrace\begin{array}{@{}l@{}}}
{\end{array}\right.}

\newcommand{\rev}[1]{{\color{red}#1}}
\newcommand{\REV}[1]{{\color{blue}#1}}
\newcommand{\revv}[1]{{\color{cyan}#1}}

\newcommand{\numberset}{\mathbb}
\newcommand{\N}{\numberset{N}}
\newcommand{\Z}{\numberset{Z}}
\newcommand{\R}{\numberset{R}}
\newcommand{\T}{\numberset{T}}
\newcommand{\C}{\numberset{C}}

\newtheorem{prop}{Proposition}
\newtheorem{thm}{Theorem}
\newtheorem{lem}[thm]{Lemma}
\theoremstyle{definition}
\newtheorem{dfn}{Definition}
\theoremstyle{remark}
\newtheorem*{rmk}{Remark}

\title[Polarized ensembles of random pure states]{Polarized ensembles of random pure states}

\author{Fabio  Deelan Cunden$^{1}$, Paolo Facchi$^{2,3}$, Giuseppe Florio$^{4,2,3}$}
\address{$^{1}$Dipartimento di Matematica, Universit\`a di Bari, I-70125 Bari, Italy}
\address{$^{2}$Dipartimento di Fisica and MECENAS, Universit\`a di Bari, I-70126 Bari, Italy} 
\address{$^{3}$INFN, Sezione di Bari, I-70126 Bari, Italy}
\address{$^{4}$Museo Storico della Fisica e Centro Studi e Ricerche
``Enrico Fermi'', Piazza del Viminale 1, I-00184 Roma, Italy}

\date{\today}

\begin{abstract}
A new family of polarized ensembles of random pure states is presented. These ensembles are obtained by linear superposition of two random pure states with suitable distributions, and are quite manageable. We will use the obtained results for two purposes: on the one hand we will be able to derive an efficient strategy for sampling states from  isopurity manifolds. On the other, we will characterize the deviation of a pure quantum state from  separability under the influence of noise.   
\end{abstract}

\pacs{03.67.Mn, 03.65.Aa, 02.50.-r 
}

\vspace{2pc}
\noindent{\it Keywords}:  Entanglement, Random matrices, Statistics
\maketitle

\section{Introduction}
In the last years many researchers  have been investigating  the typical properties of random pure states, i.e.\
unit vectors drawn at ``random'' from the Hilbert space associated to a quantum system. This subject has attracted the attention in several directions, and some important results have been achieved mostly dealing with the characterization of entanglement \cite{Zyc2,scott,Zyc3,winter2,facchi,giraud,giraud2,Facchi2,Gaussian,cappellini,majumdar,nadal,Kumar,typical}. 

The standard ensemble which has been intensively investigated is that of random pure states  with measure induced by the Haar measure on the unitary group. This ensemble, being the maximally symmetric one, implements in a natural way the  case of minimal knowledge on a quantum state \cite{Hall}. It is structureless, in the sense that the induced measure only depends on the dimension of the total Hilbert space and it is not sensible to any tensor product structure \cite{Zyc2,adachi}.

For this reason, the unitarily invariant ensemble is also known as the \emph{unbiased ensemble} \cite{typical}. A natural question is whether this ensemble of pure random states can be used to construct different, more complicated, ones. 

Various approaches have been proposed independently by several groups that have introduced different physically motivated measures on the space of pure states.  Recently, Zyczkowski \textit{et al.}~\cite{zyczk} have analyzed some kind of \emph{structured ensembles} of random pure states on composite systems that are invariant under local unitary transformations. From another perspective, De Pasquale \textit{et al.}~\cite{depasquale} have proposed a classical statistical mechanics approach in order to explore the \emph{isopurity manifolds} of random states. In the same spirit, Mueller \textit{et al.}~\cite{mueller} have recently investigated ensembles of \emph{random pure states with fixed expectation value} of some observable, in the framework of the concentration of measure phenomenon. However, there are still many obstructions in carrying out these programs, and the links among them is not yet clear.

This paper is intended as another step toward new scenarios beyond the unbiased ensemble. This step is motivated as an operational way to capture the isopurity manifolds, and turns out to be in particular cases similar to the structured ensembles proposed in~\cite{zyczk}.  Our idea is to move beyond the unbiased ensemble by using a natural operation at hand in the Hilbert space, namely  superposition of vector states. In this work we will  show that unitarily invariant measures interact nicely with the operation of superposition of states. 

As a remark, we want to stress that, in the large size limit, the robustness of  Mar\v{c}enko-Pastur's  theorem  \cite{MP} prevents many of the potentially workable ensembles to deviate from the Mar\v{c}enko-Pastur law. The approach that we propose here is a way to circumvent such an obstruction and allows to obtain workable ensembles with entanglement spectra (i.e.\  the densities of Schmidt eigenvalues) that can sensibly differ from  the Mar\v{c}enko-Pastur law. 
In this way one can investigate, by varying the strength and/or the type of the polarization, the structure of the different spectral types and the possible emergence of  phase transitions of entanglement~\cite{Facchi2}. 

The paper is organized as follows. In section \ref{sec:ensemble} we introduce the concept of polarized ensembles of pure states that will play a central role in our work. In particular, we will consider the deviation from the unbiased ensemble induced by the Haar measure on the unitary group acting on the whole Hilbert space of the system. In section \ref{sec:typicallocal} we will characterize the polarized ensembles by using the local purity of a subsystem. This approach will be used both for the study of the ensembles and for the description of an efficient procedure for sampling typical states from an isopurity manifold (and, therefore, with a fixed value of bipartite entanglement). In section \ref{sec:robust} we will apply these results to the characterization of separability of quantum states under the influence of noise. Finally, in section \ref{sec:conclusion} we will draw some conclusions. Some technical points are discussed in two appendices for completeness and self-consistency. The appendices are dedicated respectively to the computation of the average purity and to the concentration  around the average.

\section{Polarized ensembles of pure states}\label{sec:ensemble}
Consider a bipartite quantum system described by a pure state in a finite-dimensional Hilbert space $\mathcal{H}$. The bipartition in two subsystems $(A,B)$ will induce on $\mathcal{H}$ a tensor product structure such that 
\begin{equation}
\mathcal{H}=\mathcal{H}_A\otimes\mathcal{H}_B.
\end{equation} 
We will consider, without loss of generality, the case with 
\begin{equation}
\dim{\mathcal{H}_A}=N \leq \dim{\mathcal{H}_B}=M,
\end{equation} 
whence $\dim{\mathcal{H}}=NM$.

Let us focus  on the general situation in which the (pure) state of the quantum system has the form (up to normalization)
\begin{equation}
\ket{\psi}=\alpha\ket{\psi_1}+\beta\ket{\psi_2},
\label{eq:general}
\end{equation}
where $\ket{\psi_1},\ket{\psi_2}\in \mathcal{H}$ are  random states, sampled according to arbitrary probability measures.
Once the probability distributions of $\ket{\psi_1}$ and $\ket{\psi_2}$ are specified, the random variable $\ket{\psi}$
is characterized by a well defined distribution. 

Due to its privileged role in many context of random matrix theory \cite{mehta,tao}, in the following we will make large use of the unitarily invariant (unbiased) probability  measure on pure states, denoted by $\mu_{MN}$, induced by the Haar measure on the unitary group $\mathcal{U}(\mathcal{H})$. Moreover, we will consider the product measure $\mu_N\times\mu_M$ on pure states, which is left invariant under the action of local unitaries $\mathcal{U}(\mathcal{H}_A)\times\mathcal{U}(\mathcal{H}_B)$. In the latter case, we recall that the space of pure states is foliated in orbits of local unitaries labelled by the degeneracy of the Schmidt coefficients. Each of these orbits is a natural domain for the locally invariant measure $\mu_N\times\mu_M$~\cite{sinolecka,Zyc}. 

In general, depending on the ensembles chosen for  sampling  $\ket{\psi_1}$ and $\ket{\psi_2}$, the state $\ket{\psi}$ defined in~(\ref{eq:general}) will exhibit very different properties. Let us briefly outline the relevant features in some cases of particular interest. 

If both $\ket{\psi_1}$ and $\ket{\psi_2}$ are sampled according to the unbiased ensemble $\mu_{MN}$, then $\ket{\psi}$ is  also a random state whose distribution is invariant under the action of the unitary group $\mathcal{U}(\mathcal{H})$ (independently on the values of $\alpha$ and $\beta$). 

The opposite situation occurs when $\ket{\psi_1}$ is a fixed pure state and $\ket{\psi_2}\sim\mu_{NM}$. In this case the weights $\alpha$ and $\beta$ are relevant; if $\alpha\gg\beta$ the unbiased ensemble becomes ``polarized'' along the direction defined by $\ket{\psi_1}$. 
This polarization phenomenon is of particular interest if one wants to study the deviation of the properties of an ensemble of quantum states from a fixed reference state. In particular, in this work we will specialize our analysis to the cases of  $\ket{\psi_1}$ separable or maximally entangled (in the bipartite sense).

\section{Typical local purity}\label{sec:typicallocal}

\subsection{Local purity of one-parameter polarized ensembles}
In this section we will focus on the consequences of the polarization of the ensemble on the properties of bipartite entanglement between the subsystems $A$ and $B$. As a measure of entanglement we will consider the local purity of one of the parties. Given a pure state $\ket{\psi}\in\mathcal{H}$, it is defined by
\begin{equation}
\pi_{AB}(\ket{\psi})=\mathrm{Tr}_A \rho_A^2, \qquad \rho_A= \mathrm{Tr}_B \ket{\psi}\bra{\psi},
\label{eq:puritydef}
\end{equation}
where $\rho_A$ is the reduced density matrix of party $A$. The upper bound $\pi_{AB}=1$ and the lower bound $\pi_{AB}=1/\dim{\mathcal{H}_A}=1/N$ correspond, respectively, to separable and maximally entangled states.

Let us consider a state $\ket{\psi}$ obtained as a superposition~(\ref{eq:general}) where, in particular, a fixed pure state is superposed to an unbiased random one. We will get the following one-parameter ensemble
\begin{equation}
\ket{\psi}=\left[\epsilon\mathbbm{1}_{AB}+\sqrt{1-\epsilon^2}U_{AB}\right]\ket{\phi_0}\ ,
\label{eq:ensemble}
\end{equation}
where the normalized state $\ket{\phi_0}\in \mathcal{H}$ is fixed, $\epsilon\in[0,1]$ is a tunable parameter, $\mathbbm{1}_{AB}$ is the identity operator, and $U_{AB}\in\mathcal{U}\left(\mathcal{H}\right)$ is a random unitary acting on $\mathcal{H}$, sampled according to the Haar measure on the unitary group. The state 
\begin{equation}
\label{eq:phidef}
\ket{\phi}=U_{AB}\ket{\phi_0}
\end{equation} 
is therefore a (unit) random vector distributed according to the unitarily invariant measure on  pure states $\mu_{NM}$. 

Notice that for $\epsilon=0$ one recovers the unbiased ensemble. On the other hand, values of $\epsilon\ne 0$ play the role of an offset which parametrizes, as discussed in the previous section, the degree of polarization of the ensemble in the direction of $\ket{\phi_0}$. We also observe that, given two independent and symmetrically distributed random states $\ket{\phi_1},\ket{\phi_2}$, the expectation value of their overlap vanishes, i.e.
\begin{equation}
\mathbb{E}\left[\braket{\phi_1}{\phi_2}\right]=0. 
\end{equation}  
As a consequence, the normalization of the state $\ket{\psi}$ in~(\ref{eq:ensemble}) is assured on the average, in the sense that
\begin{equation}\label{eq:norm}
\mathbb{E}\left[\braket{\psi}{\psi}\right]=\mathbb{E}\big[\left\|\psi\right\|^2 \big]=1, 
\end{equation}  
and deviations from the average are exponentially suppressed for large $N$, as shown in~\ref{app:conc}. 
  
 We emphasize that, since we are focusing on the typical features of an ensemble of random pure states, any statement in the paper has to be considered in the large size limit, $N\to\infty$. In this limit, the ensemble of vectors (\ref{eq:ensemble}) is an ensemble of physical states, in the sense that it consists of unit vectors with overwhelming probability.

We are interested in the conditional expectation value of the purity $\pi_{AB}(\ket{\psi})$ given a fixed state $\ket{\phi_0}$ and a bias $\epsilon$. Due to concentration of measures, for large $N$ this quantity will be the typical purity of the polarized ensemble~(\ref{eq:ensemble}).

The density operator associated to the random pure state $\ket{\psi}$ reads
\barr
\ket{\psi}\bra{\psi}&=&\epsilon^2\ket{\phi_0}\bra{\phi_0}+(1-\epsilon^2)\ket{\phi}\bra{\phi}\nonumber\\
&+&\epsilon\sqrt{1-\epsilon^2}\left(\ket{\phi_0}\bra{\phi}+\ket{\phi}\bra{\phi_0}\right),
\earr
where $\ket{\phi}$ is given by~(\ref{eq:phidef}).
We will use the following notation:
\barr
\sigma&=&\mathrm{Tr}_B\ket{\phi}\bra{\phi},\nonumber \\
\sigma_0&=&\mathrm{Tr}_B\ket{\phi_0}\bra{\phi_0}, \nonumber\\
S_{0\phi}&=&\mathrm{Tr}_B\left(\ket{\phi_0}\bra{\phi}+\ket{\phi}\bra{\phi_0}\right).
\label{eq:3}
\earr
By tracing over subsystem $B$ and performing a straightforward calculation, we obtain 
the purity  (which is a random variable)
\barr
\fl \qquad \pi_{AB}(\ket{\psi})&=&\mathrm{Tr}_A{\rho_A^2}\nonumber\\
\fl \qquad &=&\epsilon^4\mathrm{Tr}{\sigma_0^2}+{\left(1-\epsilon^2\right)}^2\mathrm{Tr}{\sigma^2}
+\epsilon^2\left(1-\epsilon^2\right)\mathrm{Tr}{S_{0\phi}^2}
+ 2\epsilon^2\left(1-\epsilon^2\right)\mathrm{Tr}{\bigl(\sigma_0\sigma\bigr)}\nonumber\\
\fl \qquad & & + 2\epsilon^3\sqrt{1-\epsilon^2}\mathrm{Tr}{\bigl(\sigma_0S_{0\phi}\bigr)}
+2\epsilon{\left(1-\epsilon^2\right)}^{3/2}\mathrm{Tr}{\bigl(\sigma S_{0\phi}\bigr)}.
\label{eq:purity_coherent}
\earr

\subsection{Typical local purity}\label{sec:localgaussian}
We now evaluate the expectation value of the purity $\pi_{AB}$. 
The computation can be easily done by making use of a Gaussian approximation. More precisely, we will consider random vectors $\ket{\phi}\in\mathcal{H}$ whose components in an arbitrary basis are independent  complex random variables normally distributed, $\mathcal{N}_{\mathbb{C}}\left(0,1/NM\right)$ (the normalization of $\ket{\phi}$ is assured on the average). 

The Gaussian approximation is fully justified for our purposes.
Indeed, in the large size limit, concentration phenomena  and the simultaneous convergence of the Gaussian measure to the unitarily invariant measure on the sphere~\cite{Ledoux} provide the typicality of our results (see~\ref{app:conc} for further details). 
Thus, averages on the unitary group can be substituted with averages with respect to Gaussian measures.
In this case, expectation values of any smooth quantity of interest $f(\ket{\phi})$ can be easily estimated. 

We claim that the typical local purity of~(\ref{eq:ensemble}) depends only on the local purity of the pure state $\ket{\phi_0}$, 
\begin{equation}
\pi_0=\pi_{AB}(\ket{\phi_0}),
\end{equation} 
and \emph{not} on the particular  vector $\ket{\phi_0}$ with the given purity.
Indeed, a direct calculation with $\ket{\phi}$ a Gaussian vector shows (see \ref{app:explicitcalc} for explicit calculations) that the only non vanishing terms in the expectation value of~(\ref{eq:purity_coherent}) are
\begin{eqnarray}
 \label{eq:squares1} \mathbb{E}[\mathrm{Tr}{\sigma_0^2}]&=&\mathrm{Tr}{\sigma_0^2} = \pi_0,\\
 \label{eq:squares3} \mathbb{E}[\mathrm{Tr}{\sigma^2}]&=&\frac{M+N}{MN},\\
 \label{eq:squares2} \mathbb{E}[\mathrm{Tr}{\left(\sigma_0 \sigma\right)}] &=& \frac{1}{N},\\
 \label{eq:squares4} \mathbb{E}[\mathrm{Tr}{S_{0\phi}^2}]&=&\frac{2}{M}.
\end{eqnarray} 
By plugging (\ref{eq:squares1})-(\ref{eq:squares4}) into~(\ref{eq:purity_coherent}) we finally get the conditional expectation value of the purity $\pi_{AB}(\ket{\psi})$
\begin{equation}
\bar{\pi}_{AB}=\mathbb{E}[\pi_{AB}\,|\,\ket{\phi_0},\epsilon]=\epsilon^4 \pi_0+\left(1-\epsilon^4\right)\frac{M+N}{MN}\ .
\label{eq:GaussianApprox}
\end{equation}
This is a central result of the paper.

Notice that if we had performed the average using the unitarily invariant measure $\mu_{NM}$ for $\ket{\phi}$, as in~(\ref{eq:phidef}), the only difference with the above calculation would have been in the term 
\begin{equation}
\mathbb{E}[\mathrm{Tr}{\sigma^2}]=\frac{M+N}{MN+1},
\label{eq:Lub}
\end{equation} 
as computed by Lubkin \cite{lubkin}.
The relative difference with the typical purity of the unbiased Gaussian ensemble
\begin{equation}
\label{eq:piunb}
\pi_{\mathrm{unb}}=\bar{\pi}_{AB}|_{\epsilon=0}=
\frac{M+N}{MN},
\end{equation}
is thus of order $O(1/(NM))$ and negligible for large systems.

We point out an important consequence of~(\ref{eq:GaussianApprox}): 
even if $\ket{\phi_0}$ is substituted by a state $\ket{\phi'_0}=U_A\otimes U_B \ket{\phi_0}$, with $U_{A(B)}\in\mathcal{U}\left(\mathcal{H}_{A(B)}\right)$, belonging to its local orbit (therefore evolving on an isopurity manifold with arbitrary measure), the value of the typical purity given by\ (\ref{eq:GaussianApprox}) is not affected. In other words, the one-parameter ensemble of random states
\begin{equation}
\ket{\psi}=\left[\epsilon U_A\otimes U_B+\sqrt{1-\epsilon^2}U_{AB}\right]\ket{\phi_0}\ ,
\label{eq:ensemble2}
\end{equation}
where $ U_A\otimes U_B$ is a random local unitary, has average purity given by formula (\ref{eq:GaussianApprox}), with $\pi_0=\pi_{AB}(\ket{\phi_0})=\pi_{AB}(U_A\otimes U_B\ket{\phi_0})$. The ensemble (\ref{eq:ensemble2}) is a linear superposition of two random pure states with suitable probability distributions.

Incidentally, this can also be seen as a consequence of a fundamental property of  conditional expectations:
\begin{equation}
\mathbb{E}
\left[f(X,Y)\right]=\mathbb{E}
\big[\mathbb{E}
\left[f(X,Y)\,|\,Y\right]\big] ,
\end{equation} 
where $f$ is a function of two random variables $X$ and $Y$.
We will make frequent use of this property in the following.

\subsection{Typical purity for a separable and a maximally entangled polarizing state $\ket{\phi_0}$}
In this section we will discuss two interesting examples of the behavior of $\bar{\pi}_{AB}$, given  the value of $\pi_0$.
We will specialize our results about the generalized ensemble (\ref{eq:ensemble2}) to the extremal situations of a separable or maximally entangled polarizing state $\ket{\phi_0}$.

Let us start by considering the case of $\ket{\phi_0}$ separable with respect to the bipartition $(A,B)$, so that $\pi_0=1$. 
According to the discussion at the end of the Section \ref{sec:localgaussian}, we can allow $\ket{\phi_0}$ to be a random pure separable state and consider the generalized ensemble
\begin{eqnarray}
\ket{\psi}&=&\left[\epsilon U_A\otimes U_B+\sqrt{1-\epsilon^2}U_{AB}\right]\ket{\phi_{\mathrm{sep}}},\\
\ket{\phi_{\mathrm{sep}}} &=&\ket{\phi_0}_A\otimes\ket{\phi_0}_B
\label{eq:genensemblesep}
\end{eqnarray}
where $\ket{\phi_0}_A$ and $\ket{\phi_0}_B$ are fixed  states in $\mathcal{H}_A$ and $\mathcal{H}_B$,  respectively, $U_A$ and $U_B$ are random local unitaries  and $U_{AB}$ is a random global unitary transformation. The typical purity~(\ref{eq:GaussianApprox}) of the one-parameter ensemble with a separable polarization~(\ref{eq:genensemblesep}) reads
\begin{equation}
\mathbb{E}\left[\pi_{AB}\,|\,\ket{\phi_{\mathrm{sep}}},\epsilon\right]=\frac{M+N}{MN}+\epsilon^4\,\frac{MN-M-N}{MN}\ .
\label{eq:GaussianApproxSep}
\end{equation}

The other extreme case is given by $\ket{\phi_0}=\ket{\phi_{\mathrm{ent}}}$, a maximally entangled pure state, so that $\pi_0=1/N$. Then we are dealing with a polarized ensemble of the form 
\begin{equation}
\ket{\psi}=\left[\epsilon U_A\otimes U_B+\sqrt{1-\epsilon^2}U_{AB}\right]\ket{\phi_{\mathrm{ent}}}\ ,
\label{eq:genensembleme}
\end{equation}
where the reference state is such that 
\begin{equation}
\mathrm{Tr}_B{\ket{\phi_{\mathrm{ent}}}\bra{\phi_{\mathrm{ent}}}}=\frac{1}{N}\mathbbm{1}_A. 
\end{equation}
Such a polarization decreases the typical purity 
of the unbiased ensemble to the value
\begin{equation}
\mathbb{E}\left[\pi_{AB}\,|\,\ket{\phi_{\mathrm{ent}}},\epsilon\right]=\frac{M+N}{MN}-\frac{\epsilon^4}{M}\ .
\label{eq:GaussianApproxEnt}
\end{equation}
These results will be compared in the following section to a numerical Monte Carlo approach.

\subsection{Generation of random pure states with fixed local purity}
The results obtained in the previous sections suggest a very inexpensive strategy for generating random pure states with fixed value of the purity $\pi_{AB}$.
Indeed, the numerical sampling of the uniform measure on the manifold of pure states $\ket{\psi}$ parametrized by a fixed value of $\pi_{AB}$ will proceed through the following steps:
\begin{enumerate}
\item Choose $\epsilon\in[0,1]$ such that 
\begin{equation}\label{eq:pi0choice}
\pi_{AB}=\epsilon^4 \pi_0+\left(1-\epsilon^4\right) \pi_{\mathrm{unb}},
\end{equation}
where $\pi_0=1$ or $\pi_0=1/N$ if the desired value of $\pi_{AB}$ is, respectively, larger or smaller than the unbiased typical value $ \pi_{\mathrm{unb}}$ in~(\ref{eq:piunb}).
\item Generate a pure state $\ket{\psi}$ by superposition
\begin{equation}
\ket{\psi}=\epsilon\ket{\phi_0}+\sqrt{1-\epsilon^2}\ket{\phi}\ ,
\end{equation}	
where $\ket{\phi}$ is sampled according to the unbiased measure $\mu_{NM}$ and $\ket{\phi_0}$ is a  separable or maximally entangled pure state sampled randomly according to the invariant measure under local unitaries $\mu_N\times\mu_M$, depending on the value of $\pi_0$ chosen in~(\ref{eq:pi0choice}).
\end{enumerate}

\begin{figure}[t]
\centering
\includegraphics[width=.9\columnwidth]{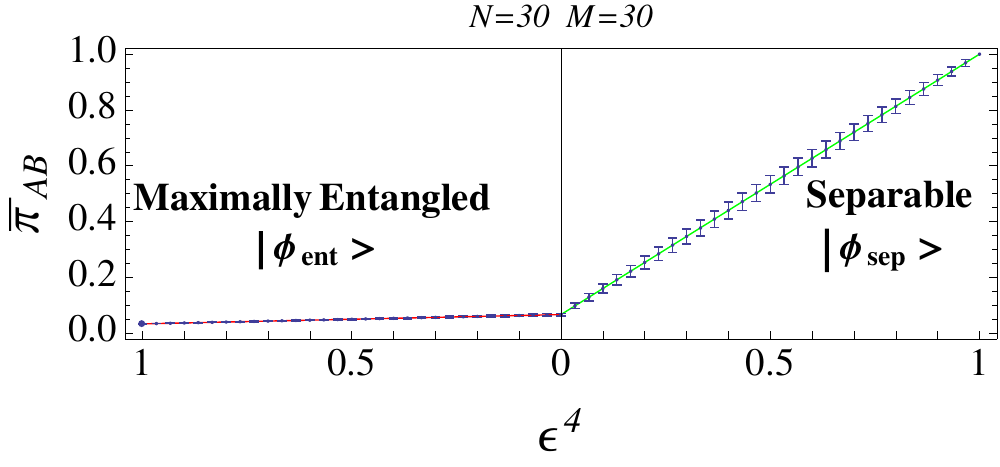}\vspace{.5cm}
\includegraphics[width=.9\columnwidth]{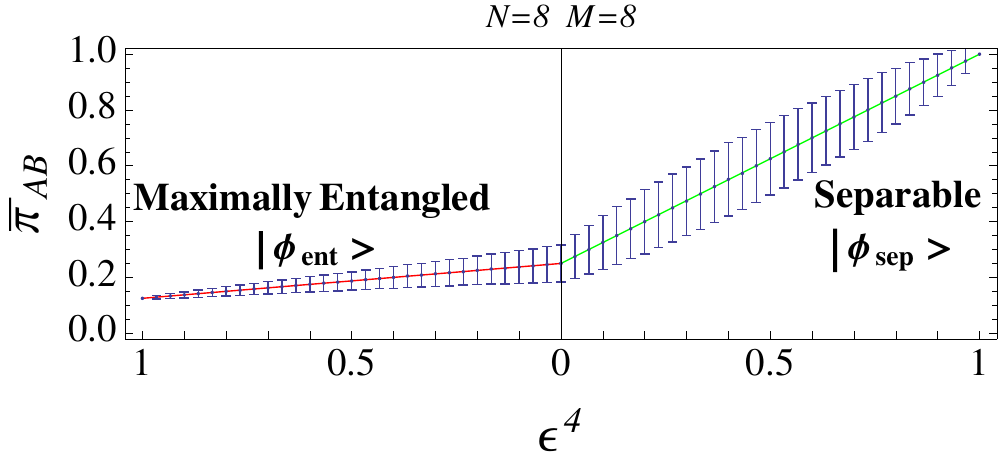}
\caption{(Color online) Typical purity of polarized ensembles vs $\epsilon^4$, 
depending on the bias  $\ket{\phi_0}$. We compare the analytical prediction (continuous lines) with the numerical values of $\pi_{AB}$ (sample mean and error bars) obtained from the sampling procedure described in the  text. We  consider balanced bipartitions with size $N=M=30$ (top) and  $N=M=8$ (bottom). In both cases the number of realizations used to perform the ensemble average is $n=10^4$.
Right side: the continuous line represents the analytical prediction for the purity of an ensemble polarized by a separable state~\ (\ref{eq:GaussianApproxSep}). Depending on the value of the parameter $\epsilon$ the purity ranges from the unbiased value $\pi_{\mathrm{unb}}=(M+N)/MN$ to the maximum $\pi_0=1$. Left side: the continuous line represents the analytical prediction for the purity of an ensemble polarized by a maximally entangled state~\ (\ref{eq:GaussianApproxEnt}). As  $\epsilon$ increases the typical value of the purity decreases from $\pi_{\mathrm{unb}}$ to the minimum $\pi_0=1/N$.
The error bars represent the standard deviations of the numerical simulations from the estimated average. Such fluctuations are exponentially suppressed as the dimensions $N,M$ increase,  according to Eq.~(\ref{eq:GaussApplied}) in~\ref{app:conc}. } 
\label{fig:GaussExp}
\end{figure}

In figure \ref{fig:GaussExp} (upper panel) the analytical formulas~(\ref{eq:GaussianApproxSep}) and (\ref{eq:GaussianApproxEnt}) are compared to the Monte Carlo results for the values of $\bar{\pi}_{AB}$ obtained by sampling  pure states through the procedure outlined above. The comparison  shows clearly the efficiency of the sampling procedure in providing  the correct behavior of the typical purity vs the bias $\epsilon$ with quite small fluctuations already for dimensions $N=M=30$. Such fluctuations around the average are more evident for smaller systems. See figure \ref{fig:GaussExp} (lower panel) for the case  $N=M=8$.

\section{Robustness of separability under random perturbations}\label{sec:robust}
As an application, in this section we will use the results obtained from the study of polarized ensembles  to analyze the stability of  ``separability'' of quantum states with respect to random additive perturbations. More precisely, if the state of the system $\ket{\phi_0}\in\mathcal{H}_A\otimes\mathcal{H}_B$ is separable, $\ket{\phi_0}=\ket{\xi_0}_A\otimes\ket{\chi_0}_B$, how much noise is necessary to make the reduced state $\rho_A=\mathrm{Tr}_B{\left[\ket{\phi_0}\bra{\phi_0}\right]}$ distinguishable from a pure state? 
We notice that the problem of the characterization of separability has attracted a lot of interest in the context of the analysis of ground states of quantum spin systems \cite{illuminati3,Rezai,Rossini}.

Let us consider the state of the bipartite system in the form
\begin{equation}
\ket{\psi}=\sqrt{1-\eta^2}\ket{\xi_0}_A\otimes\ket{\chi_0}_{B}+\eta\ket{\phi}\ ,\quad 0\leq\eta\leq1,
\label{eq:setup}
\end{equation}
where $\ket{\phi}\sim\mu_{NM}$ is an unbiased  random perturbation, and $\eta$ measures the strength of the  noise. (Notice that in the language of  polarized ensembles of the previous sections we  have $\epsilon=\sqrt{1-\eta^2}$). 

In the context of spin systems, the parameter $\eta$ should be related e.g.\ to the magnetic field that drives the ground state from a separable state to an entangled one. 
More generally, this setup models the following situation: some non-controllable noise prevents one to prepare with infinite precision a given state. The noise has the consequence that the reduced density matrix $\rho_A$ is not a projection but is mixed. It is reasonable that, for weak noise ($\eta\ll 1$), we still obtain a reduced state $\rho_A$ that is close to  a projection. Then the question is: is there  a threshold on the value of noise above which separability is appreciably destroyed? 

In order to answer this question, it is convenient to introduce the notion of \emph{effective dimension} (also known as  ``participation ratio'' \cite{Bell}) of a state $\rho$ defined as 
\begin{equation}
d^{\mathrm{eff}}(\rho)=\frac{1}{\mathrm{Tr}{\rho^2}}\in\left[1,N\right].
\label{eq:EffDim}
\end{equation}
The effective dimension of a mixed state quantifies how many pure states appreciably contribute to the mixture. Moreover, differently from the rank of $\rho$, $d^{\mathrm{eff}}$ captures the probabilistic weight of different states and is more manageable for explicit calculations.

For a separable pure state, $\ket{\xi_0}_A\otimes \ket{\chi_0 }_B$,  the reduced density matrix, $\rho_A=\ket{\xi_0}\bra{\xi_0}$, has effective dimension
$d^{\mathrm{eff}}(\rho_A)=\mathrm{rank}(\rho_A)=1$.
A global perturbation acting on $\ket{\xi_0}_A\otimes \ket{\chi_0 }_B$ can be appreciated locally if the reduced state $\rho_A$ becomes a mixture. In order to obtain a mixture one needs at least two pure components. Therefore, we can say that a quantum state is distinguishable from a one-dimensional projection if its effective dimension $d^{\mathrm{eff}}$ is equal or larger than $2$. 
As a consequence, we are led to the following criterion on the separability of  the  state averaged over the noise realization: 
\begin{equation}\label{eq:constraint}
d^{\mathrm{eff}}(\mathbb{E}[\rho_A])<2.
\end{equation}
We get that
\begin{equation}
d^{\mathrm{eff}}(\mathbb{E}[\rho_A])
=\frac{1}{\mathbb{E}[\mathrm{Tr}{\rho_A^2}]}=\frac{1}{\mathbb{E}\big[\pi_{AB}\,|\,\ket{\phi_{\mathrm{sep}}},\sqrt{1-\eta^2}\big]}\ ,
\end{equation}
and then, from~(\ref{eq:GaussianApproxSep}), it is straightforward to prove that  condition~(\ref{eq:constraint}) is satisfied if
\begin{equation}
\eta^2<\eta_{\star}^2(\pi_{\mathrm{unb}})=1-\sqrt{\frac{1-2\pi_{\mathrm{unb}}}{2-2\pi_{\mathrm{unb}}}},
\end{equation}
where $\pi_{\mathrm{unb}}$ is the typical purity of the unbiased ensemble~(\ref{eq:piunb}).
If $\eta\geq\eta_{\star}$ the local state $\rho_A$ will be mixed with high probability. 

In the limit of large system sizes, $N,M\rightarrow+\infty$, the threshold critical value becomes
\begin{equation}
\eta_{\star}^2=\left(1-\frac{1}{\sqrt{2}}\right)+O\left(\frac{1}{N}\right),
\end{equation}
since $1/N\leq \pi_{\mathrm{unb}}\leq 2/N$.
Therefore, as long as the state $\ket{\psi}$ of the large quantum system has the form (\ref{eq:setup}) with
\begin{equation}
\eta<\sqrt{1-\frac{1}{\sqrt{2}}}\simeq0.54,
\end{equation}
one has $d^{\mathrm{eff}}(\rho_A)<2$, and separability will be (approximately) preserved. Notice that in this case, by applying perturbation theory, one gets that the spectrum of $\rho_A$ is made of a large eigenvalue of order $O(1-\eta^2)$ and a sea of eigenvalues of order $O(1/N)$ that have a negligible influence on the reduced density matrix $\rho_A$.

\section{Conclusions}\label{sec:conclusion}
In this paper we have shown that, using the superposition principle, we can take advantage of the knowledge of the unbiased ensemble of random pure states in order to explore new interesting ensembles. In particular, we have found that  adding a bias in a suitable direction is enough to polarize the unitarily invariant ensemble.  We stress that our approach has been oriented to the study of  typical bipartite entanglement between subsystems, as measured by the local purity. 

This strategy yields an efficient and simple sampling of random pure states with fixed value of purity, and paves the way to further explorations and a deeper characterization of the geometry of isopurity manifolds. 

Finally, we have applied our results to the analysis of  separability of quantum states under the influence of random perturbation modeled through a coherent superposition. For large systems, we have obtained a critical value of the noise strength, independent of the system size, beyond which the state is no longer separable, and the reduced state gets appreciably  mixed.

\ack
This work is partially supported by PRIN 2010LLKJBX.
The authors acknowledge support from the University of
Bari through the Project IDEA. GF acknowledges support
by INDAM through the Project Giovani GNFM.

\appendix

\section{Evaluation of  $\bar{\pi}_{AB}$} \label{app:explicitcalc} 
In this section we detail the derivation of the average purity $\bar{\pi}_{AB}$ given in~(\ref{eq:GaussianApprox}).
As mentioned in the text, the averages  are more easily obtained by switching from the unitarily invariant measure to the Gaussian measure. 
The random state in (\ref{eq:ensemble}) reads
\begin{equation}
\ket{\psi}=\epsilon\ket{\phi_0}+\sqrt{1-\epsilon^2}\ket{\phi}\ ,
\label{eq:ensemble_app}
\end{equation}
where $\ket{\phi}$ is a Gaussian random vector and   $\ket{\phi_0}$   a reference vector. Let the $NM$ complex components of $\ket{\phi}$ and $\ket{\phi_0}$ (in a given basis) be $X_{i\mu}$ and $A_{i\mu}$, respectively, where $1\leq i\leq N$ and $1\leq \mu\leq M$. In the following the $X_{i\mu}$'s will be independent and identically distributed (iid) complex Gaussian variables  with mean $\mathbb{E}\left[X_{i\mu}\right]=0$ and variance $\mathbb{E}\left[|X_{i\mu}|^2\right]=1/NM$. 
The one-dimensional projection $\ket{\phi}\bra{\phi}$ has random entries $X_{i\mu}X^*_{j\nu}$, where the star denotes  complex conjugation. The reduced state of subsystem $A$, obtained by partial trace, $\sigma=\mathrm{Tr}_B\ket{\phi}\bra{\phi}$ is an $N$-dimensional square random matrix whose entries are $\sum_\mu{X_{i\mu}X^*_{j\mu}}$. We are interested in objects like 
\begin{equation}
\mathrm{Tr}{\sigma^2}=\sum_{i,j,\mu,\nu}{X_{i\mu}X^*_{j\mu}X_{j\nu}X_{i\nu}^*}.
\end{equation}
 Recall that the only ingredient necessary to deal with a collection of iid complex Gaussian  random variables  is
the Wick formula for the expectation of the products
\begin{equation}
\mathbb{E}[X_{i_1}\cdots X_{i_{n}} X_{j_1}^*\cdots X_{j_{n}}^*]=\sum_p {\mathbb{E}[X_{i_1} X^*_{j_{p(1)}}] \cdots \mathbb{E}[X_{i_{n}} X^*_{j_{p(n)}}]}
\label{eq:Wick}
\end{equation}
where the sum is over all  possible permutation $p$ of $n$ elements, and 
\begin{equation}
\mathbb{E}[X_i X^*_j]=(NM)^{-1} \delta_{ij}.
\end{equation}
The expectations of all other products vanish.
Thus, the average value of the purity of state $\ket{\phi}$  reads
\begin{eqnarray}
\mathbb{E}[\mathrm{Tr}{\sigma^2}]&=&\mathbb{E}[\sum_{i,j,\mu,\nu}{X_{i\mu}X^*_{j\mu}X_{j\nu}X^*_{i\nu}}]\nonumber\\
&=&\sum_{i,j,\mu,\nu}\left\{{\mathbb{E}[X_{i\mu}X^*_{j\mu}]\,\mathbb{E}[X_{j\nu}X^*_{i\nu}]
+\mathbb{E}[X_{i\mu}X^*_{i\nu}]\,\mathbb{E}[X_{j\nu}X^*_{j\mu}]}\right\}
\nonumber\\
&=&\frac{1}{\left(MN\right)^2}
\sum_{i,j,\mu,\nu}{\left(\delta_{i j}+\delta_{\mu\nu}\right)}
=\frac{1}{\left(MN\right)^2}
\left(M^2 N+M N^2\right)
\nonumber \\
&=&\frac{M+N}{MN}\ .
\label{eq:computation1}
\end{eqnarray}
Let us consider now the terms (\ref{eq:squares2})-(\ref{eq:squares4}). The reduced state $\sigma_0=\mathrm{Tr}_B\ket{\phi_0}\bra{\phi_0}$  
has components $\sum_\mu{A_{i\mu}A^*_{j\mu}}$. Then 
\begin{eqnarray}
\mathbb{E}[\mathrm{Tr}{\left(\sigma_0\sigma\right)}]&=&\mathbb{E}[\sum_{i,j,\mu,\nu}{A_{i\mu}A^*_{j\mu}X_{j\nu}X^*_{i\nu}}]
=\sum_{i,j,\mu,\nu}{A_{i\mu}A^*_{j\mu}\mathbb{E}[X_{j\nu}X^*_{i\nu}]}\nonumber\\
&=&\frac{1}{MN}\sum_{i,\mu,\nu}{|A_{i\mu}|^2}=\frac{1}{MN}\sum_{\nu}{1}=\frac{1}{N}\ .
\label{eq:computation2}
\end{eqnarray}
The last term $S_{0\phi}=\mathrm{Tr}_B\left(\ket{\phi_0}\bra{\phi}+\ket{\phi}\bra{\phi_0}\right)$ is a random matrix with entries $\sum_\mu{A_{i\mu}X^*_{j\mu}+X_{i\mu}A^*_{j\mu}}$. By squaring it and taking the trace we obtain
\begin{eqnarray}
\mathbb{E}[\mathrm{Tr}{S_{0\phi}^2}]&=&
\mathbb{E}[\sum_{i,j,\mu,\nu}{
\left(A_{i\mu}X^*_{j\mu}+X_{i\mu}A^*_{j\mu}\right)
\left(A_{j\nu}X^*_{i\nu}+X_{j\nu}A^*_{i\nu}\right)}]\nonumber\\
&=&2\sum_{i,j,\mu}{|A_{i\mu}|^2\mathbb{E}[|X_{j\mu}|^2]}\nonumber\\
&=&\frac{2}{MN}\sum_{i,j,\mu}{|A_{i\mu}|^2}=\frac{1}{MN}\sum_{j}{1}=\frac{2}{M}\ .
\label{eq:computation3}
\end{eqnarray}
Adding up all the pieces we obtain the result (\ref{eq:GaussianApprox}). 

Finally, we notice that the computation with a Gaussian measure deviates from the computation with a uniform measure on the sphere only in the four-point correlation~(\ref{eq:computation1}). Indeed for a unit vector $\ket{\phi}$  uniformly distributed on the unit sphere the Wick theorem~(\ref{eq:Wick}) is no longer valid  and the fourth moment is slightly modified into
\begin{equation}
\mathbb{E}[X_{i\mu}X^*_{j\mu}X_{j\nu}X^*_{i\nu}]
=\frac{1}{MN(MN+1)}
\left(\delta_{i j}+\delta_{\mu\nu}\right) ,
\label{eq:sphere_mom}
\end{equation}
(see equation~(53) of \cite{cumulants}) which gives (\ref{eq:Lub}) in place of~(\ref{eq:computation1}).

\section{Gaussian approximation and typicality}
\label{app:conc} 

The typicality of the average purity in the polarized ensembles used in Section \ref{sec:localgaussian} relies on the following concentration phenomenon for Gaussian variables~\cite{tao}:
\begin{lem}
Let $\bm{X}=\left(X_1,X_2,\dots,X_k\right)$ be a vector with independent identically distributed Gaussian components, with distribution $X_i\sim\mathcal{N}(0,\sigma^2)$.
Then, for any smooth function $f: \numberset{R}^k\rightarrow\numberset{R}$, with $\eta=\sup{|\nabla f|}<\infty$, the following concentration inequality holds
\begin{equation}
\mathbf{Pr}\left\{\left| f(\bm{X})-\mathbb{E}[f] \right| >\alpha\right\}\leq 2\exp{\left(-\frac{\alpha^2}{4\eta^2 \sigma^2}\right)}.
\end{equation}
\end{lem}

Let us now consider the polarized ensemble defined in (\ref{eq:ensemble}).
In the Gaussian approximation the $2MN$ real coordinates of the random vector state are Gaussian i.i.d. random variables with distribution $\mathcal{N}\left(0,1/2MN\right)$, so that normalization is assured on average. 
However, since $\|\cdot\|$ has Lipschitz constant 1, $\|\psi\|$ has $\eta\leq \sqrt{1-\epsilon^2}$ and thus   the ensemble~(\ref{eq:ensemble}) is composed of normalized vectors with overwhelming probability:
\begin{equation}
\mathbf{Pr}\left\{\left| \|\psi\|^2 - 1 \right| >\alpha\right\}\leq 2\exp{\left(-\frac{NM \alpha^2}{2(1-\epsilon^2)}\right)}.
\end{equation}

Moreover, the local purity function $\pi_{AB}$, defined in~(\ref{eq:puritydef}), has Lipschitz constant bounded by $\eta\leq 4$ (see \cite{winter}). The Lipschitz constant of the purity of the polarized ensemble (\ref{eq:ensemble}) is thus bounded by $\eta\leq 4\sqrt{1-\epsilon^2}$ and then from the Lemma one gets
\begin{equation}
\mathbf{Pr}\Bigl\{ \Bigl|\pi_{AB}(\ket{\psi})-\mathbb{E}[\pi_{AB}]\Bigr| >\alpha\Bigr\}\leq 2\exp{\left(-\frac{NM\alpha^2}{32(1-\epsilon^2)}\right)} .
\label{eq:GaussApplied}
\end{equation}

Incidentally, we  mention that a similar Gaussian tail can be derived for uniformly distributed unit vectors by Levy's lemma \cite{winter}. The proof relies on a judicious use of $\delta$-nets. In a finite dimensional setting, any subset of the sphere of states is totally bounded, in the sense that it admits a finite $\delta$-net, for all $\delta>0$. What we are interested in is a bound on the cardinality of a $\delta$-net $\mathcal{N}$  on manifolds of equal Schmidt rank $k$. For such manifolds a bound is given by 
\begin{equation}
|\mathcal{N}|\leq\left(\frac{10}{\delta}\right)^{2k(N+M)}.
\label{eq:est}
\end{equation}
For a proof see \cite{winter}. In this work the authors bound the cardinality of $\delta$-nets on set with fixed Schmidt rank. In fact, what they obtain  are $\delta$-nets on orbits of pure states under local unitaries, i.e. states with fixed Schmidt coefficients.
Estimate~(\ref{eq:est}) is good enough to control the probability of deviations of the purity from its average, so that a bound of the form~(\ref{eq:GaussApplied}) is obtained.


\section*{References}
\begin{thebibliography}{99}


\bibitem{Zyc2}
\.{Z}yczkowski K and  Sommers H-J  2001
{\it J. Phys. A} \textbf{34} 7111

\bibitem{scott}
Scott A J and Caves C M 2003 \JPA \textbf{36} 9553

\bibitem{Zyc3}
Sommers H-J and \.{Z}yczkowski K  2004
\JPA \textbf{37} 8457

\bibitem{winter2}
Hayden P, Leung D W, Shor P W and Winter A 2004
{\it Commun. Math. Phys.} \textbf{250} 371

\bibitem{facchi}
Facchi P, Florio G and Pascazio S 2006
\PR A \textbf{74} 042331

\bibitem{giraud}
Giraud O 2007
{\it J. Phys. A: Math. Theor.} \textbf{40} 2793 

\bibitem{giraud2}
Giraud O 2007
{\it J. Phys. A: Math. Theor.} \textbf{40} F1053


\bibitem{Facchi2} 
Facchi P, Marzolino U, Parisi G, Pascazio S, and Scardicchio A 2008
\PRL \textbf{101} 050502 


\bibitem{Gaussian}
Lupo C, Mancini S, De Pasquale A, Facchi P, Florio G and Pascazio S 2012
\JMP \textbf{53} 122209 


\bibitem{cappellini}
Cappellini V, Sommers H-J and \.{Z}yczkowski K 2006
\PR A \textbf{74} 062322 

\bibitem{majumdar}
Majumdar S N, Bohigas O and Lakshminarayan A 2008
{\it J. Stat. Phys.} \textbf{131} 33 

\bibitem{nadal}
Nadal C, Majumdar S N and Vergassola M 2010
\PRL \textbf{104} 110501 

\bibitem{Kumar}
Kumar S and Pandey A 2011
{\it J. Phys. A: Math. Theor.} \textbf{44} 445301 

\bibitem{typical}
Cunden F D, Facchi P, Florio G and Pascazio S 2013
\textit{Eur. Phys. J. Plus} \textbf{128} 48

\bibitem{Hall}
Hall M J W 1998
\PL A \textbf{242} 123

\bibitem{adachi}
Adachi S, Toda M and Kubotani H 2009
{\it Annals of Physics} \textbf{324}  2278

\bibitem{zyczk}
\.{Z}yczkowski K, Penson K A, Nechita I and Collins B 2011 
\JMP \textbf{52} 062201  

\bibitem{depasquale}
De Pasquale A, Facchi P, Parisi G, Pascazio S and Scardicchio A 2010
\PR A \textbf{81} 052324

\bibitem{mueller}
Mueller M, Gross D and Eisert J 2011 
{\it Comm. Math. Phys.}  \textbf{303} 785 

\bibitem{MP}
Mar\v{c}enko V A and Pastur L A1967 
{\it Math. USSR-Sb.} \textbf{1} 457-483 

\bibitem{mehta}
Mehta M L 2004
\textit{Random Matrices} (San Diego: Elsevier AcademicPress) 

\bibitem{tao}
Tao T 2012 
\textit{Topics in Random Matrix Theory} (American Mathematical Society) 

\bibitem{sinolecka}
Sino\l\c{e}cka M M, \.{Z}yczkowski K and Ku\'{s} M 2002 Acta Physica Polonica B \textbf{33} 2081 

\bibitem{Zyc}
Bengtsson I and \.{Z}yczkowski K 2006 
\textit{Geometry of quantum states} (Cambridge University Press)

\bibitem{Ledoux}
Ledoux M 2005 
\textit{The concentration of measure phenomenon} (American Mathematical Society) 

\bibitem{lubkin}
Lubkin E 1978 
\JMP \textbf{19} 1028 

\bibitem{illuminati3}
Giampaolo S M, Adesso G and Illuminati F 2009 
\PR B \textbf{79} 224434

\bibitem{Rezai}
Rezai M, Langari A and Abouie J 2010 
\PR B \textbf{81} 060401

\bibitem{Rossini}
Tomasello B, Rossini D, Hamma A and  L. Amico 2011
{\it Europhysics Letters} \textbf{96} 27002

\bibitem{Bell}
Bell R J and  Dean P 1970
{\it Discuss. Faraday Soc.} \textbf{50} 55

\bibitem{cumulants}
P. Facchi P, Florio G, Marzolino U, Parisi G and Pascazio S 2010
\textit{J. Phys. A: Math. Theor.} \textbf{43} 225303

\bibitem{winter}
Hayden P, Leung D W and Winter A 2006 
{\it Comm. Math. Phys.} \textbf{265} 95


\end{thebibliography}
\end{document}